\documentclass[final,5p, times,twocolumn,showpacs]{revtex4}
 \usepackage{verbatim} 
\usepackage{tikz}
\usetikzlibrary{shapes.geometric, calc} 
\usepackage{lineno,hyperref}
\usepackage{graphicx}
\usepackage{ epstopdf }
\usepackage{amsmath}
\usepackage[vcentermath]{youngtab}

\begin{document}


\title{Compact pentaquark structures} 

\author{Elena Santopinto, Alessandro Giachino} 
\affiliation{INFN Sezione di Genova,  via Dodecaneso 33, 16164 Genova, Italy}

\begin{abstract}
\noindent	
We study the possibility that at least one of the two  pentaquark structures recently 
reported by  $LHC_b$  \cite{Aaij:2015tga,Aaij:2016phn,Aaij:2016ymb} could be  
described as a compact pentaquark state, and we give predictions  for new channels 
that can be studied by the experimentalists if this hypothesis is correct. 
We use  very general  arguments dictated by symmetry considerations, in order 
to describe the pentaquark  states within a group theory approach. 
A complete classification of all possible states and quantum numbers, which can be 
useful both to the 
experimentalists in their search for new findings and to theoretical model builders, 
is given, without  
the introduction of any particular dynamical model.
 Some predictions are finally given by means of a  G\"ursey-Radicati (GR) inspired mass 
 formula. 
  We reproduce the mass and the quantum numbers of the lightest pentaquark state  
 reported by $LHC_b$ 
 ($J^{P}=\frac{3}{2}^{-}$) with a parameter-free  mass formula, fixed on the 
 well-established baryons.
We predict other pentaquark resonances (giving their masses, and suggesting  
possible decay channels) 
which belong to the same multiplet as the lightest one. 
Finally, we compute the partial decay widths for all the predicted 
pentaquark resonances.
\end{abstract}


\maketitle


\section{Introduction}
\begin{figure}[h!]
	\centering
	\includegraphics[width=0.6\linewidth]{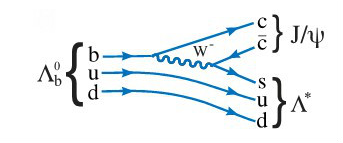}
	\caption{Feynman diagrams for (\ref{decay1}) $\Lambda_{b}^{0}\rightarrow J/\psi+\Lambda^{*}$  
	(Fig. taken from Ref. \cite{Aaij:2015tga}; APS copyright) }
	\label{fig:feyn1}
\end{figure}

\begin{figure}[h!]
	\centering
	\includegraphics[width=0.6\linewidth]{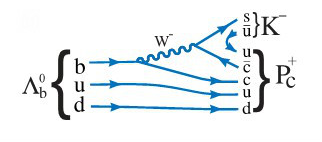}
	\caption{Feynman diagrams for (\ref{decay2}) $\Lambda_{b}^{0}\rightarrow P^{+}_{c}+K^{-}$  
	(Fig. taken from Ref. \cite{Aaij:2015tga}; APS copyright)}
	\label{fig:feyn2}
\end{figure}

The $LHC_b$ collaboration has recently reported the observation of exotic structures in $\Lambda_b$ 
decay \cite{Aaij:2015tga}, further supported by another two articles of the $LHC_b$ 
collaboration \cite{Aaij:2016phn,Aaij:2016ymb}.  \\The decay can proceed according to the diagram 
in fig. \ref{fig:feyn1}, which involves conventional hadrons:
\begin{equation}
\Lambda_{b}^{0}\longrightarrow J/\psi+\Lambda^{*}
\label{decay1}
\end{equation}
or it can be  characterised by exotic contributions, which  are referred to as charmonium-pentaquark 
states (fig. \ref{fig:feyn2}):
\begin{equation}
\centering
\Lambda_{b}^{0}\longrightarrow P^{+}_{c}+K^{-}
\label{decay2}{\large }
\end{equation}
The $LHC_b$ collaboration found two resonant structures: a lower mass state at  $4380\pm8\pm28\; MeV$, 
with a width of  $205 \pm 18\pm 86\; MeV$, and a higher mass one at $4449.8\pm 1.7 \pm 2.5\; MeV$, with 
a width of $39\pm5\pm19\; MeV$, in the $J/\Psi p$ invariant mass spectrum. 
Moreover, according to the $LHC_b$ collaboration \cite{Aaij:2015tga}, the preferred  $J^{P}$ assignments 
are $3/2^{-}$ and $5/2^{+}$, respectively.
\\\\
Since the observation of the two resonant structures, many explanations have been proposed for 
the $LHC_b$ pentaquark states.
Meson-baryon molecules were suggested in 
\cite{Karliner:2015ina,Chen:2015loa,Roca:2015dva,He:2015cea,Meissner:2015mza,Chen:2015moa}.
Pentaquark states of diquark-diquark-antiquark nature were suggested in \cite{Maiani:2015vwa,Wang:2015epa}, and $\bar{D}$ soliton states in \cite{Scoccola:2015nia}.\\ 
The molecular interpretation works  well  for the heaviest resonant state (see, for example 
\cite{Karliner:2015ina}). 
Therefore, in this study, we focus on the lightest pentaquark structure (with $J^{P}=\frac{3}{2}^{-}$), 
by means of a multiquark approach.
From the $LHC_b$ quantum numbers of the lightest resonant state, we show that 
it can be described as 
a pentaquark state with spin $S=\frac{3}{2}$.
We  show that the ground state multiplet of the charmonium pentaquark states 
is a $SU_{f}(3)$ octet, and we studied all the  charmonium pentaquark states which 
belong to the octet, predicted their masses, and
suggested possible decay channels in which the experimentalists can observe them. 
By using an effective Lagrangian \cite{Kim:2011rm} for 
the $P_{c} J/\Psi$ coupling,
in combination with the branching ratio $\mathcal{B}(P_c^+ \to J/\Psi p)$ upper  
limit extracted by Wang \cite{Wang:2015jsa}, and with our predicted masses, 
we compute the partial
decay widths for the predicted pentaquark resonances.


\section{ Classification of the {\lowercase{ \fontsize{0.45cm}{1pt}\selectfont 
$\mathbf{ qqqc\bar{c} }$}} multiplets as based 
on symmetry properties}	
In order to classify the pentaquark multiplets, we made use, as much as possible, 
of symmetry principles,
without introducing any explicit dynamical models.
We  made use of the Young tableaux technique, adopting for each representation the 
notation 
$[f]_d=[f_1,\ldots,f_n]_d$, where $f_i$ denotes the number of boxes in the $i$-th row 
of the Young tableau,
and $d$ is the dimension of the representation.\\ \\
In agreement with the $LHC_b$ hypothesis \cite{Aaij:2015tga},
we think the charmonium pentaquark wave function as 
$qqqc\bar{c}$  where $q=u,d,s$ is a light quark  and $c$ is the heavy charm  quark.
Let us first discuss the possible configurations of $qqq$ quarks
in the $qqqc\bar{c}$ system.\\
\\The $c\bar{c}$ can be in a colour octet or singlet with spin 0 or 1.
The colour wave function of the $qqqc\bar{c}$ system must be 
an $SU_{c}(3)$ singlet, so 
the remaining three light quarks are also in a  color-singlet or in a color-octet.
\\
\indent The orbital symmetry of the quark wave function depends on the 
quantum numbers of the resonant state
$J^{P}=\frac{3}{2}^{-}$.
Indeed, the  parity $P$ of the pentaquark system is:
\begin{equation}
P\mid qqqc\bar c>=(-1)^{l+1}
\label{parity}
\end{equation}
and so $l$ must be even.
The total angular momentum is $J=\frac{3}{2}$, and so $l=0$ or $l=2$. In this paper, we  hypothesise that the lightest charmonium pentaquark state 
reported by the $LHC_b$ collaboration  is a ground state pentaquark with $l=0$,
and so each quark is in $S$-wave.
\\
\indent The  three  light quarks must satisfy the Pauli principle.
As a consequence, since the $q^3$ orbital part is completely symmetric,
the spin-flavour   and the colour part  
are conjugate:  spin-flavour  symmetric state if they are in a colour singlet;
or spin-flavour mixed symmetry state if they are in a colour octet.
Therefore, the  allowed $SU_{sf}(6)$ spin-flavour  pentaquark configurations are
a 56-plet  ($[3]_{56}$)  and a 70-plet  ($[21]_{70}$) which correspond respectively
to the three light quarks in a colour singlet and in a colour octet.
In  Tab. \ref{sfqqq} the analysis of the flavour and spin content of the spin-flavour
56-plet and of the 70-plet , i.e. their decomposition into 
the representations of $SU_f (3) \otimes SU_s (2)$, is reported.
\begin{table}[h!]
\centering
\caption[]{Spin-flavour decomposition of the two allowed $SU_{sf}(6)$ spin-flavour 
pentaquark configurations: the  56-plet   and the 70-plet.}
\vspace{15pt}
\label{sfqqq}
\begin{tabular}{llclcl}
\hline
 & & & & \\
 $SU_{\rm sf}(6)$ & $\supset$ & $SU_{\rm f}(3)$ 
& $\otimes$ & $SU_{\rm s}(2)$ \\
 & & & & \\
\hline
 & & & & \\
 $[3]_{56}$ & & $[3]_{10}$  & $\otimes$ & $[3]_{4}$ \\
                 & & $[21]_{8}$ & $\otimes$ & $[21]_{2}$ \\
 & & & & \\
  $[21]_{70}$ & & $[3]_{10}$  & $\otimes$ & $[21]_{2}$ \\
                & & $[21]_{8}$ & $\otimes$ & $[3]_{4}$ \\
                & & $[21]_{8}$ & $\otimes$ & $[21]_{2}$ \\
               & & $[111]_{1}$ & $\otimes$ & $[21]_{2}$ \\

& & & & & \\
\hline
\end{tabular}
\end{table}
\\
The $SU_{sf}(6)$ 70-plet  contains
an  $SU_{f}(3)$ flavour octet $[21]_{8}$ and a decuplet $[3]_{10}$,
while the 56-plet contains an  $SU_{f}(3)$ flavour singlet $[111]_{1}$,
two octets $[21]_{8}$ and a decuplet $[3]_{10}$.
The allowed $SU_{f}(3)$ flavour representations to which the charmonium
pentaquark states can belong are therefore:
\begin{equation}
[111]_{1}\;,\;\;[21]_{8} \;,\;\;[3]_{10} 
\label{allowed multi1}
\end{equation}

Since the charmonium pentaquark state, as reported by $LHC_b$, has a quark
content $uudc\bar{c}$, it does not have strange quarks, and so 
the strangeness ${\bf S} =0$ .
In the case of $3$ flavours ($u,d,s$), the hypercharge $Y$ is defined as:
\begin{equation}
Y=B+{\bf S}
\label{hypercharge}
\end{equation}
where $B$ is the barionic number and ${\bf S}$ is the strangeness.
Since the charmonium pentaquark state, as reported
by LHCb, has a quark content $uudc\bar{c}$, it does not have
strange quarks, and so the strangeness ${\bf S}=0$, 
the charm $C$ is 0, the barionic number $B=1$, and then $Y$ must be equal to 1.
\\
Therefore the pentaquark state  must be found in a $Y = 1$ submultiplet of the 
allowed flavour states of  Eq. \ref{allowed multi1}.
Following this reasoning, we must exclude  the singlet $[111]_{1}$, because it 
does not have any $Y=1$ submultiplets; 
therefore,  the remaining  possible $SU_{f}(3)$ multiplets for the charmonium
pentaquark states are:
\begin{equation}
[21]_{8} \;,\;\;[3]_{10} 
\label{allowed multi2}
\end{equation}
\section{The extension of the G\"ursey-Radicati mass formula}
In order to determine the mass splitting between the multiplets  
of Eq. \ref{allowed multi2},
we  made use of a G\"ursey-Radicati (GR)-inspired formula \cite{F.Gursey}.
As yet, there is experimental evidence of only two charmonium pentaquark states.
This is not sufficient to determine all parameters in the GR mass formula, and 
then to predict the masses of 
the other pentaquarks. For this reason, we use the values of the parameters 
determined from the three-quark spectrum 
(see Tab. \ref{baryons used}), assuming that the coefficients in the GR formula 
are the same for different quark systems.  
The simplest GR formula extension  which permits us 
to distinguish the different multiplets of $SU_{f}(3)$ is 
\begin{eqnarray}
M_{GR}=M_0+AS(S+1)+DY+ \nonumber\\  + E\left[I(I+1)-\frac{1}{4}Y^2\right] +GC_2(SU(3))+FN_C
\label{GR}
\end{eqnarray} 
where $M_0$ is a scale parameter: this means that, for example, in baryons each quark gives  a contribution of 
roughly $\frac{1}{3}M_0$ to the whole mass.\\
$I$ and $Y$ are the isospin and hypercharge, respectively, while $C_2(SU(3))$ is 
the eigenvalue of the $SU_{f}(3)$ 
Casimir operator.
Finally,  $N_C$ is a counter of $c$ quarks or $\bar{c}$ antiquarks. 
This term takes into account the mass difference 
between a $c$ quark (or a $\bar{c}$ antiquark) in relation to the light quarks ($u,d$).
The coefficients $A,D,E,G,F$ and the scale parameter $M_{0}$ have been fixed by using 
the well-established  baryons spectrum.
\\
Tab. \ref{baryons used} reports the baryons  used  to fix the parameters in Eq. \ref{GR} , the $SU_{f}(3)$  
multiplet  which they  were assigned to,  the corresponding eigenvalues of 
the Casimir operator $C_2(SU(3))$, their quantum numbers, 
and the values of $N_c$. \\
In Tab. \ref{parameters} all the parameters, with their corresponding values, 
are reported.
\begin{table}[h!]
	\begin{tabular}{ccccccc}
		\hline
		&&\\
		baryon & $SU_{f}(3)$  & $C_2(SU(3))$ & spin & $Y$ & $I$ & $N_c$  \\
		& multiplet&&&&&\\
		$\Lambda(1116)$ & $[21]_{8}$ & $3$ & $\frac{1}{2}$ & $0$ & 0 & 0   \\
		$\Lambda_{C}^{+}(2286)$ & $[11]_{3}$ & $\frac{4}{3}$ & $\frac{1}{2}$ & $\frac{2}{3}$ & 0 & 1   \\
		$\Sigma_{C}^{+}(2455)$ & $[2]_{6}$ & $\frac{10}{3}$ & $\frac{1}{2}$ & $\frac{2}{3}$ & 1 & 1  \\
		$\Xi_{C}^{+}(2471)$ & $[11]_{3}$ & $\frac{4}{3}$ & $\frac{1}{2}$ & $-\frac{1}{3}$ & $\frac{1}{2}$  & 1  \\
		$\Xi_{C}^{+'}(2576)$ & $[2]_{6}$ & $\frac{10}{3}$ & $\frac{1}{2}$ & $-\frac{1}{3}$ & $\frac{1}{2}$  & 1  \\
		$\Omega_{C}^{0}(2695)$ & $[2]_{6}$ & $\frac{10}{3}$ & $\frac{1}{2}$ & $-\frac{4}{3}$ & 0 & 1  \\
		$\Omega_{C}^{+}(2766)$ & $[2]_{6}$ & $\frac{10}{3}$ & $\frac{3}{2}$ & $-\frac{4}{3}$ & 0  & 1  \\
	\end{tabular}
	\caption{On the left side, the list of baryons used to fix the parameters in Eq. \ref{GR}, 
	is reported. 
	In the following columns we report the multiplet   which they were assigned 
	to, with the corresponding 
	eigenvalues of the Casimir operator  $C_2(SU(3))$, their quantum numbers, 
	and the values of $N_c$, respectively. 
	}
		\label{baryons used}
\end{table}
\\
\begin{table}[h!]
	\begin{tabular}{ccccccc}
		\hline
		&&&&&&\\
		& $M_{0}$ & $A$ & $D$ & $E$ & $F$ & $G$ \\
		&&&&&\\
		\hline
		&&&&&\\
		value ($MeV$) & 940,8 & 23,6 & -157,3 & 32,0 & 1365,7 & 52,5 \\
	\end{tabular}
	\caption{values of the parameters  in the GR extended mass formula \ref{GR}}
	\label{parameters}
\end{table}
\\\\
In order to show the reliability of the values obtained with the GR mass formula 
extension, we calculated 
the predicted mass of the two charmed baryons $\Sigma_{c}(2520)$ and $\Xi_{c}(2645)$ 
reported by the PDG \cite{PDG} .
 The quantum number assignments and the 
 predicted masses are reported 
  in Tabs. \ref{bar mass prediction0} and  \ref{bar mass prediction},
 respectively.
 \begin{table}[h!]
 \begin{tabular}{ccccccc}
		\hline
		&&&&&&\\
		baryon & $SU_{f}(3)$  & $C_2(SU(3))$ & spin & $Y$ & $I$ & $N_c$ \\
		& multiplet&&&&& \\
		&&&&&&\\
	$\Sigma_{C}$ & $[2]_{6}$ & $\frac{10}{3}$ & $\frac{3}{2}$ & $\frac{2}{3}$ & 1 & 1  \\
		&&&&&&\\
		$\Xi_{C}$ & $[2]_{6}$ & $\frac{10}{3}$ & $\frac{3}{2}$ & $-\frac{1}{3}$ & $\frac{1}{2}$  & 1\\
	\end{tabular}
	\caption{On the left side, the  $\Sigma_{C}$ and $\Xi_{C}$ baryons. On the right side, 
	the $SU_{f}(3)$ multiplet  
	which they were assigned to, with  eigenvalues of the Casimir operator $C_2(SU(3))$, 
	the assigned quantum numbers, 	and the values of $N_c$.}
	\label{bar mass prediction0}
	\end{table}
\begin{table}[h!]
	\begin{tabular}{cccc}
		\hline
		&&&\\
		baryon & isospin& exp. masses  & predicted masses \\
		&configurations&($MeV$) & ($MeV$)\\
		&&&\\                                         
		$\Sigma_{C}(2520)$ & 	$\Sigma_{C}^{++}$& $2518.41^{+ 0.21}_{-0.19}$ &2526 \\ 
		&&&\\
			 & 	$\Sigma_{C}^{+}$& $2517.5\pm 2.3$ &2526 \\   
				&&&\\
		 & 	$\Sigma_{C}^{0}$& $2518.48\pm 0.20$ &2526 \\ 
				&&&\\
			$\Xi_{C}(2645) $& $\Xi_{C}^{+}$ & $2645.9\pm 0.5$ & 2646\\ 
			&&&\\
		& $\Xi_{C}^{0}$ & $2645.9\pm 0.5$ & 2646\\
	\end{tabular}
	\caption{The first column shows  $\Sigma_{C}$ and $\Xi_{C}$ baryons; the possible isospin configurations 
	are reported in the second column, while in the third column the corresponding experimental 
	masses as from  PDG \cite{PDG}.  In the last column the predicted masses, calculated by means of the mass
	formula of Eq. \ref{GR} 
using the coefficients of Tab. \ref{parameters} , are reported.
	}
	\label{bar mass prediction}
	\end{table}


\section{Application of the GR formula to the pentaquark states}
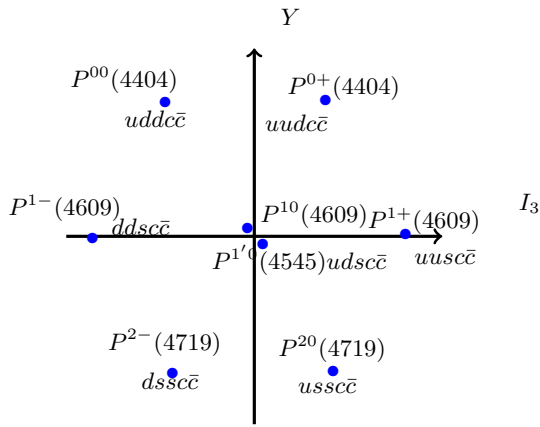
\begin{figure}[h]
  		\begin{tikzpicture}[scale=2.5]
  		\centering
  	\draw [very thick,->] (-1,0)--(1,0);
  	\draw [very thick,->] (0,-1)--(0,1);
  		
  		\node (pol) [ thick, blue!90!black,minimum size=4cm,regular polygon,rotate=117.5, regular polygon sides=6] at (0,0) {}; %
  		
  		\foreach \n [count=\nu from 0, remember=\n as \lastn, evaluate={\nu+\lastn}] in {1,2,...,5,6} 
  		\node[anchor=\n*(360/6)] at(pol.corner \n) {\color{blue}\circle*{4}}; %
  		\node[anchor=280] at(0,0) {\color{blue}\circle*{4}};
  		\node[anchor=150] at(0,0) {\color{blue}\circle*{4}};   		
  		\node[anchor=340] at(pol.corner 6) {$P^{00}(4404)$};
  		\node[anchor=390] at(pol.corner 6) {$uddc\bar{c}$};
  		\node[anchor=250] at(pol.corner 5) {$P^{0+}(4404)$};
  		\node[anchor=380] at(pol.corner 5) {$uudc\bar{c}$};
  		\node[anchor=230] at(pol.corner 4) {$P^{1+}(4609)$};
  		\node[anchor=160] at(pol.corner 4) {$uusc\bar{c}$};
  		\node[anchor=200] at(0,0) {$P^{10}(4609)$};
  		\node[anchor=152] at(0,0) {$P^{1'0}(4545)${$udsc\bar{c}$}};
  		\node[anchor=330] at(pol.corner 1) {$P^{1-}(4609)$};
  		\node[anchor=185] at(pol.corner 1) {$ddsc\bar{c}$};
  		\node[anchor=250] at(pol.corner 3) {$P^{20}(4719)$};
  		\node[anchor=100] at(pol.corner 3) {$ussc\bar{c}$};
  		\node[anchor=290] at(pol.corner 2) {$P^{2-}(4719)$};
  		\node[anchor=80] at(pol.corner 2) {$dssc\bar{c}$};
  		\put(100,10){$I_3$}
  		\put(10,80){$Y$}
  		\end{tikzpicture}
  		\caption{octet of the charmonium pentaquark states: each state is 
  		labelled with $P^{ij}(M)$, where $i=0,1,2$ is the number of strange 
  		quarks of a given pentaquark state,  $j=-,0,+$ is the pentaquark's 
  		electric charge, and $M$  the predicted mass.}         			
  		\label{fig:octetedited}
  	\end{figure}

In Eq.  \ref{allowed multi2} we reported the possible  $SU_{f}(3)$  multiplets for the charmonium pentaquark states.
We  hypothesise that the charmonium pentaquark state  $J^P=\frac{3}{2}^{-}$, reported by the $LHC_b$ collaboration,	
belongs to the lowest mass $SU_{f}(3)$ multiplet.
According to the GR formula \ref{GR}, the mass splitting between the different  $SU_{f}(3)$ multiplets of Eq.  
\ref{allowed multi2} is due to the different eigenvalues of the  Casimir operator $C_2(SU(3))$, and so it is proportional 
to the coefficient $G$ (reported in Tab. \ref{parameters}).
Since $G$ is positive ($G=52,5\;MeV$), the lowest mass multiplet is the one with the minimum  Casimir operator eigenvalue, 
and so it is the octet (see Tab.  \ref{casimir charmonium penta states}).
In Tab. \ref{casimir charmonium penta states}, each multiplet, with the corresponding eigenvalues of 
the Casimir operator $C_2(SU(3))$, is reported.
\begin{table}[h!]
	\begin{tabular}{cc}
		\hline
		&\\
		$SU_{f}(3)$ multiplet &  $C_2(SU(3))$  \\
		&\\
		\hline
		&\\
		$[3]_{10}$ & 6 \\
		$[21]_{8}$ & 3 \\
		&\\
		\hline
	\end{tabular}
	\caption{The possible charmonium pentaquark multiplets (eq. \ref{allowed multi2}), with their corresponding 
	eigenvalues of the Casimir operator  $C_2(SU(3))$, is reported. }
	\label{casimir charmonium penta states}
\end{table}
\\From Tab. \ref{casimir charmonium penta states}, we can see that the lowest mass 
charmonium pentaquark state 
is the $[21]_{8}$ $SU_{fl}(3)$ octet.
Therefore, in this octet, we  expect to find the charmonium pentaquark state $J^P=\frac{3}{2}^{-}$ reported by the $LHC_b$ 
collaboration.
In the following,  we focus on the octet charmonium pentaquark states, and we apply the  GR mass formula \ref{GR}, 
with the values of the parameters reported in Tab. \ref{parameters}, to each state of the octet, in order to predict 
the corresponding mass.
 As regards the notation, we indicate a charmonium pentaquark state ($qqqc\bar{c}$, with $q=u,d,s$)  
by $P^{ij}(M)$, where $i=0,1,2$ is the number of strange quarks of a given pentaquark state,  $j=-,0,+$ 
is the pentaquark's electric charge, and $M$  the predicted mass.
The state identified with the one reported by the $LHC_b$ collaboration ($P^{0+}(4404)$), and the other predicted 
charmonium pentaquark states of the octet, are reported in Fig. \ref{fig:octetedited} .
We observe that the charge state $P^{0+}(4404)$ has just the same quantum numbers as the lightest resonance 
(charge, spin, parity) reported by the $LHC_b$ collaboration.

Its theoretical mass, predicted by means of our GR formula extension, is $M=\;4404\; MeV$. \\
Despite the simplicity of the approach that we used, this result is in agreement with the mass reported by the $LHC_b$ 
collaboration: $M = 4380 \pm 8 \pm 29 \;MeV$.\\
Our compact pentaquark approach predicts that it is a member of an isospin doublet, with hypercharge $Y=1$.\\
 If the compact pentaquark description is correct, also the other octet states should be found by the $LHC_b$ collaboration. 
 On the contrary, if the $LHC_b$ pentaquark is mainly a molecular state, it is not necessary that all the states of that multiplet exist.
\section{Decay channels}
\begin{figure}[h]
	\label{P0}
	\centering
	\includegraphics[width=0.8\linewidth]{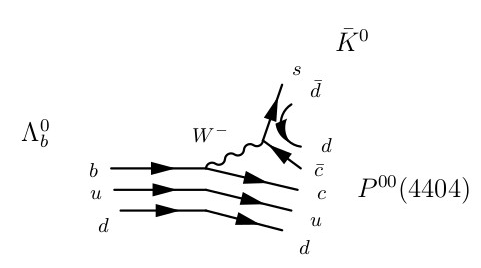}
	\caption{$\Lambda_b$  baryon decay in $P^{00}(4404)$ and $\bar{K}^{00}$, where $P^{00}(4404)$ is the neutral pentaquark 
	state, a member of the isospin doublet  with $Y=1$. }
\end{figure}

We will now explore the possible decay channels in which the other predicted states of the octet can be observed.
These channels will be described in detail.  
The state $P^{0+}(4404)$  is a part of an isospin doublet. 
In order to observe its isospin partner ($P^{00}(4404)$), a possible decay channel could be:
\begin{equation}
\Lambda_{b}^{0}\longrightarrow P^{00}+\bar{K}^{00}, P^{00} \longrightarrow J/\Psi+n \;.
\end{equation} 
The corresponding Feynman diagram is reported in Fig. \ref{P0} .

With respect to the other charmonium pentaquark states  of the octet, with strangeness,  we have to focus on the decays of bottom baryons with strange quarks.
Let us consider the following $\Xi_{b}^{-}$ decay:

\begin{equation}
\Xi_{b}^{-}\longrightarrow J/\psi + \Xi^{-} \;.
\label{Xi_b decay}
\end{equation}     
This decay is present in nature and  was discovered  by the $D0$ collaboration (\cite{V.M.Abazov_Xi_b}).
In analogy with the exotic $\Lambda_{b}^{0}$ decay of Fig. \ref{decay2} , we can expect that also in
the case of $\Xi_{b}^{-}$ baryon there is another possible exotic decay channel:
\begin{equation}
\Xi_{b}^{-}\longrightarrow  P^{10}/P^{1'0} + K^{-}, P^{10}/P^{1'0}\longrightarrow J/\Psi+\Sigma/\Lambda  \;,
\label{Xi1}
\end{equation}
where $P^{10}(4609)$ and $P^{1'0}(4535)$ have the same quark content ($usdc\bar{c}$), and belong to the isospin 
triplet, and to the isosinglet, respectively (see Fig. \ref{fig:octetedited}).
Since they have the same quark content and both are neutral, they can both come from the $\Xi_{b}^{-}$ decay.
The charmonium  pentaquark state $P^{1-}(4609)$ can be observed in the following decay process:
\begin{equation}
\Xi_{b}^{-}\longrightarrow  P^{1-} + \bar{K}^{0}, P^{1-}\longrightarrow J/\Psi + \Sigma^{-} \;.
\label{Xi2}
\end{equation}
The difference between the two suggested decays for the $\Xi_b^{-}$ baryon (Eq. \ref{Xi1}, and Eq. \ref{Xi2}) 
is in the final state: in the case of the final state of  Eq. \ref{Xi1} , 
a couple of quarks $u\bar{u}$ comes from the vacuum, while, in the decay of Eq. \ref{Xi2} , 
the couple of quarks $u\bar{u}$  is replaced with a couple $d\bar{d}$.\\
The baryon $\Xi_{b}^{-}$ is a member of an isodoublet.
The decay of its isospin partner $\Xi_{b}^{0}$ 
\begin{equation}
\Xi_{b}^{0}\longrightarrow  P^{1+} + K^{-}, P^{1+}\longrightarrow J/\Psi + \Sigma^{+}
\end{equation}
is probably the most important one from the experimental point of view, since all the final state particles  
are charged and, therefore, easier to detect.

In order to have a final pentaquark state with two strange quarks $s$, we need a double strange baryon 
in the initial state.   
The known decay channel of the $\Omega_{b}$ baryon is: 
\begin{equation}
\Omega_{b}^{-}  \longrightarrow J/\psi + \Omega^{-} \;.
\end{equation}
This decay was discovered by the $D0$ detector at the Fermilab Tevatron collider \cite{V.M.Abazov_Omega_b}.
Another possible $\Omega_{b}^{-}$  decay channel may be, in analogy with the exotic $\Lambda_{b}$  decay 
channel of Fig. \ref{decay2} :
\begin{equation}
\Omega_{b}^{-}  \longrightarrow P^{20} + K^{-}, P^{20}\longrightarrow J/\Psi + \Xi^{0}  \; .
\label{Omega1}
\end{equation}  
The state $P^{20}(4719)$  of Eq. \ref{Omega1} is a part of an isospin doublet (see Fig. \ref{fig:octetedited}). 
In order to observe its isospin partner ($P^{2-}(4719)$), a possible decay channel could be:
\begin{equation}
\Omega_{b}^{-}  \longrightarrow P^{2-} + \bar{K}^{0}, P^{2-}\longrightarrow J/\Psi + \Xi^{-}   \; .
\label{Omega2}
\end{equation}
The difference between the $\Omega_{b}^{-}$  decays of Eq. \ref{Omega1} and that of Eq. \ref{Omega2} 
is, respectively, the creation of a couple $u\bar{u}$ and $d\bar{d}$ from the vacuum.
\section{Partial decay widths}
We adopt the effective Lagrangian 
for the $P_c N J/\psi$  
couplings from Ref.  \cite{Kim:2011rm} as follows:
\begin{eqnarray}
\begin{gathered}
\mathcal{L}_{P_c N \psi}^{3/2^{-}}                                    
= i\overline {P_c}_\mu \left[ 
\frac{g_1}{2M_N} \Gamma_\nu^{-} N 
\right]\psi^{\mu\nu} +
\end{gathered}
\label{eq:ResLagrangian}
\end{eqnarray}
\begin{eqnarray}
\begin{gathered}
-i\overline {P_c}_\mu\left[\frac{ig_2}{(2M_N)^2} \Gamma^{-} \partial_\nu N {+} 
\frac{ig_3}{(2M_N)^2} \Gamma^{-} N \partial_{\nu} 
\right] \psi^{\mu\nu} + \mathrm{H.c.} \nonumber \\                              
\end{gathered}
\end{eqnarray}
where $P_{c}$ is the pentaquark field with spin-parity $J^{P}=\frac{3}{2}^{-}$,
$N$ and $\psi$ are the nucleon and the $J/\Psi$ fields, respectively. 
The $\Gamma$ matrices are
defined as follows:
\begin{align}
\Gamma_{\nu}^{-} = \left(
\begin{array}{c} 
\gamma_{\nu}\gamma_5 \\ \gamma_{\nu}
\end{array} \right)\;,
\Gamma^{-} = \left(
\begin{array}{c} 
\gamma_5 \\ \mathbf{1}
\end{array} \right) \;.
\end{align}
As noticed by Wang \cite{Wang:2015jsa} in the pentaquark state 
decays into $J/\psi p$, the momentum of the final states are fairly small compared with
the nucleon mass.
Thus, the higher partial wave terms proportional to $(p/M_{N})^{2}$ and $(p/M_{N})^{3}$
can be neglected, so we only consider  the first term in Eq.~(\ref{eq:ResLagrangian}).
This approximation leads to the following expression for the  $P_c^{0+}(4380)$ 
partial decay width in the $N J/\psi$ channel \cite{Oh:2011}: 
\begin{eqnarray}
\begin{gathered}
\Gamma (P_c^{0+} \to N J/\psi) =
\frac{{\bar{g}_{N J/\Psi}}^2}{12\pi} \frac{p_N}{M_{P_c^{0+}}} (E_N + M_N)         \\  
\times
[2E_N(E_N - M_N) + (M_{P_c^{0+}} - M_N)^2 + 2M_{J/\psi}^2]         
\end{gathered}
\label{eq:DW2}
\end{eqnarray}
with
\begin{equation}
 \bar{g}_{N J/\Psi}=\frac{g_{1}}{2M_{N}}
 \label{eq:gbar}
\end{equation}
The kinematic variables $E_N$ and $p_N$ in Eq.~(\ref{eq:DW2}) are
defined as $E_N = (M_{P_c}^2 + M_N^2 - M_{J/\psi}^2) / (2M_{P_c})$
and $p_N = \sqrt{E_N^2 -M_N^2}$.\\
\indent Unfortunately, the branching ratio  $\mathcal{B}(P_c^+ \to J/\Psi p)$ is 
not known at present, so the coupling constant $g_{1}$ of Eq. \ref{eq:gbar} is unknown.
However, by using our pentaquark mass predictions,
we can provide an expression of the partial decay widths for the 
pentaquark states with open strangeness.
For example, the $P_c^{1+} $ partial decay width in  the $\Sigma^{+} J/\Psi$ channel 
is given by:
\begin{eqnarray}
\begin{gathered}
\Gamma (P_c^{1+} \to \Sigma^{+} J/\psi) =
\frac{{\bar{g}_{\Sigma^{+} J/\Psi}}^2}{12\pi} \frac{p_{\Sigma^{+}}}{M_{P_c^{1'}}} 
(E_{\Sigma^{+}} 
+ M_{\Sigma^{+}})      \\   
\times
[2E_{\Sigma^{+}}(E_{\Sigma^{+}} - M_{\Sigma^{+}}) + (M_{P_c^{1+}} - M_{\Sigma^{+}})^2 + 
2M_{J/\psi}^2],               
\end{gathered}
\label{eq:DW3}
\end{eqnarray}
and the coupling constant $\bar{g}_{\Sigma^{+} J/\Psi}$ is:
\begin{equation}
 \bar{g}_{\Sigma^{+} J/\Psi}=\frac{g_{1}}{2M_{\Sigma^{+}}}\;.
\end{equation}
\\
The expressions for the partial decay widths  
of the $\Lambda J/\Psi$, $\Sigma J/\Psi$, and $\Xi J/\Psi$ channels  
are listed in Table~\ref{TABLE1}.
\begin{table}[thp]
\begin{tabular}{ccc} \hline\hline
initial state& channel & partial width ($MeV$) \\
$P_c^{1'0} $& $\Lambda J/\Psi$ &  $(0.81 \Gamma_{NJ/\Psi})$\\
$P_c^{1-},P_c^{10},P_c^{1+}, $& $\Sigma J/\Psi$ & $(0.73 \Gamma_{NJ/\Psi})$\\
$P_c^{2-},P_c^{20}, $& $\Xi J/\Psi$ & $(0.65 \Gamma_{NJ/\Psi})$\\
&&\\
\hline\hline
\end{tabular}
\caption{Partial decay widths expressions for $\Lambda J/\Psi$, $\Sigma J/\Psi$
and $\Xi J/\Psi$ channels.} 
\label{TABLE1}
\end{table}
\\
Since the pentaquark states were
observed in  $J/\Psi p$ channel, it is natural to expect
that they can be produced in $J/\Psi p$ photoproduction via the s
and u-channel process.
Wang et al. ~\cite{Wang:2015jsa} calculated
the pentaquark states cross section in $J/\Psi $ photoproduction and 
compared it with the present experimental data (\cite{Camerini}, \cite{Anderson},
\cite{Gittelman}). 
The coupling between  $J/\Psi p $ and  the two pentaquark states are extracted
by assuming it accounts for their total width and $5\%$, respectively.
As a result, they found that if one assumes that the $J/\Psi p$ channel  saturates 
the total width of the two
pentaquark states (that is $\mathcal{B}(P_c^+ \to J/\Psi p)=1$ )  
one significantly overestimates  the experimental data.
In conclusion they found that  to be consistent also with the present 
photoproduction data,  it is necessary that the branching ratio for both 
the pentaquark states is $\mathcal{B}(P_c^+ \to J/\Psi p) \leq 0.05$.
Thus, if we use  the upper  branching ratio limit extracted by Wang 
~\cite{Wang:2015jsa},  that is $\mathcal{B}(P_c^+ \to J/\Psi p) =0.05$, 
we obtain that the $P_{c}(4380)$  partial decay width for the $J/\psi p$  channel is 
\begin{equation}
\Gamma_{NJ/\Psi}=\mathcal{B}(P_c^+ \to J/\Psi p)\Gamma_{tot}= 10.25 \; MeV
\end{equation}
where $\Gamma_{tot}$ as reported by the $LHC_{b}$ collaboration, is $205 \;MeV$.
The numerical results for the other channels are listed 
in Table~\ref{TABLE2}.
\begin{table}[thp]
\begin{tabular}{ccc} \hline\hline
initial state& channel & partial width ($MeV$)\\
$P_c^{1'0} $& $\Lambda J/\Psi$ & $8.35$ \\
$P_c^{1-},P_c^{10},P_c^{1+}, $& $\Sigma J/\Psi$ & $7.59$ \\
$P_c^{2-},P_c^{20}, $& $\Xi J/\Psi$ & $6.69$ \\
&&\\
\hline\hline
\end{tabular}
\caption{Partial decay widths for $\Lambda J/\Psi$, $\Sigma J/\Psi$
and $\Xi J/\Psi$ channels. 
The partial decay widths are calculated from the constraint that 
$ J/\Psi p$ channel accounts  for the $5\%$ of the total pentaquark width,
as  calculated by Wang in (\cite{Wang:2015jsa}). } 
\label{TABLE2}
\end{table}
\section{Conclusions}
The $LHC_b$ collaboration has recently reported the observation of two exotic structures in $J/\Psi\;p$ 
channel \cite{Aaij:2015tga}, which they referred to as  charmonium pentaquark states 
( with a  quark content $uudc\bar{c}$ ) 
further supported by another two articles by the $LHC_b$ collaboration \cite{Aaij:2016phn,Aaij:2016ymb}. 
The significance of each of these  states is more than 9 standard  deviations. The lightest one has a 
mass of $4380 \pm 8 \pm 29\; MeV$ and a width of $ 205 \pm 18 \pm 86 \;MeV$, while the heaviest has a mass 
of $4449.8 \pm 1.7 \pm 2.5\; MeV$ and a width of $39 \pm 5 \pm 19\; MeV$.
The preferred  $J^{P}$ assignments, according to the $LHC_b$ collaboration \cite{Aaij:2015tga}, are $3/2^{-}$ and $5/2^{+}$, 
respectively.
\\
The earliest prediction for the charmonium pentaquark with $J^{P}= \frac{3}{2}^{-}$ was given by J. J. Wu et al. \cite{Wu}.
The heaviest pentaquark state has been apparently well explained by means of a molecular approach \cite{Roca:2015dva,Karliner:2015ina},
and it was also predicted in a molecular approach before the $LHC_b$ discovery by  \cite{Xiao:2013yca}, 
in a  coupled-channel 
unitary approach.
\\As regards the lightest one, molecular models have also been proposed, 
but the predictions are not so good as 
for the heaviest state \cite{Roca:2015dva,Karliner:2015ina}. 
Some predictions of its mass and quantum numbers were given  by \cite{Yuan:2012wz} 
in 2012, by means of a   
potential quark model approach, but these predictions depend strongly on the 
particular interaction used: 
colour-magnetic interaction (CM) based on one-gluon exchange, chiral interaction 
(FS ) based on meson exchange, 
and instanton-induced interaction (Inst.) based on the non-perturbative QCD vacuum 
structure.
\\In this present study, we focused on describing the lightest resonant state 
($J^P=\frac{3}{2}^{-}$), 
by means of a multiquark approach.
  An extension of the original GR mass formula \cite{F.Gursey} which  correctly 
  describes the charmed baryon 
  sector was performed  (Tab. 	\ref{bar mass prediction}), and also proved able 
  to give  an  unexpected prediction  
  for the mass of  the  lightest pentaquark state $\frac{3}{2}^{-}$, which is in 
  agreement with the experimental 
  value within one standard deviation.
\\We found that the lightest pentaquark state $\frac{3}{2}^{-}$  belonged to the  
$SU_{f}(3)$ octet $[21]_{8}$. 
The  theoretical mass of the lightest pentaquark state $\frac{3}{2}^{-}$
predicted by means of the GR formula extension (Eq. \ref{GR}) is $M=\;4404\; MeV$, 
in agreement with 
the experimental mass $M = 4380 \pm 8 \pm 29 \;MeV$.  We also  predicted  other 
pentaquark states, which  belong to 
the same $SU_{f}(3)$ multiplet as the lightest resonance $J^{P}=\frac{3}{2}^{-}$,
giving their mass, and  suggesting 
possible decay channels in which they can be observed.
We have finally computed the 
partial decay  widths for all the suggested octet-pentaquark  decay channels.
\\
\indent As the $\Lambda_{b}\longrightarrow J/\Psi K^- p$ decay 
is expected to be dominated by $\Lambda^{*}\longrightarrow K^- p $ resonances
\cite{Aaij:2015tga}, we observe that
the poor knowledge about the $\Lambda^*$  excited  states can 
affect the estimation of the parameters of the two pentaquark 
resonances.
Moreover, as was noticed by Wang ~\cite{Wang:2015jsa}, 
if the two pentaquark candidates are genuine states,
their production in photoproduction should be a natural expectation.
\indent
For these reasons, on the one hand it is important to increase our knowledge about
the missing excited  states $\Lambda^*$  with new experiments 
(\cite{proceedings}), in order 
to improve the analysis and to extract with more precision the two pentaquark
masses and widths.
On the other hand, a refined measurement of the $J/\Psi$ photoproduction
cross section would provide more information about the 
nature of the pentaquark states.



\clearpage
\begin{thebibliography}{99} 
	
	
	
	\bibitem{Aaij:2015tga}
	R.~Aaij {\it et al.} [LHCb Collaboration],
	Phys.\ Rev.\ Lett.\  {\bf 115} (2015) 072001
	
	\bibitem{Aaij:2016phn}
	R.~Aaij {\it et al.} [LHCb Collaboration],
	Phys.\ Rev.\ Lett.\  {\bf 117} (2016) no.8,  082002
	
	\bibitem{Aaij:2016ymb}
	R.~Aaij {\it et al.} [LHCb Collaboration],
	Phys.\ Rev.\ Lett.\  {\bf 117} (2016) no.8,  082003
	
	\bibitem{Karliner:2015ina}
	M.~Karliner and J.~L.~Rosner,
	Phys.\ Rev.\ Lett.\  {\bf 115} (2015) 12,  122001
	
	
	
	
	
	
	
	
	
	
	
	\bibitem{Chen:2015loa}
	R.~Chen, X.~Liu, X.~Q.~Li and S.~L.~Zhu,
	Phys.\ Rev.\ Lett.\  {\bf 115} (2015) no.13,  132002
	
	\bibitem{Roca:2015dva}
	L.~Roca, J.~Nieves and E.~Oset,
	Phys.\ Rev.\ D {\bf 92} (2015) no.9,  094003
	
	\bibitem{He:2015cea}
	J.~He,
	Phys.\ Lett.\ B {\bf 753} (2016) 547
	
	
	
	\bibitem{Meissner:2015mza}
	 U.~G.~Meissner and J.~A.~Oller,
	Phys.\ Lett.\ B {\bf 751} (2015) 59
	
	
	
	\bibitem{Chen:2015moa}
	H.~X.~Chen, W.~Chen, X.~Liu, T.~G.~Steele and S.~L.~Zhu,
	Phys.\ Rev.\ Lett.\  {\bf 115} (2015) no.17,  172001
	
	\bibitem{Xiao:2015fia}
	C.~W.~Xiao and U.-G.~Meißner,
	Phys.\ Rev.\ D {\bf 92} (2015) no.11,  114002
	
	\bibitem{Maiani:2015vwa}
	L.~Maiani, A.~D.~Polosa and V.~Riquer,
	Phys.\ Lett.\ B {\bf 749} (2015) 289
	
	\bibitem{Wang:2015epa}
	Z.~G.~Wang,
	Eur.\ Phys.\ J.\ C {\bf 76} (2016) no.2,  70
	
	\bibitem{Wu}
	J. J. Wu et al., Phys. Rev. Lett. 105 (2010) 232001
	\bibitem{Xiao:2013yca}
	C.~W.~Xiao, J.~Nieves and E.~Oset,
	Phys.\ Rev.\ D {\bf 88} (2013) 056012
	
	
	
	\bibitem{Scoccola:2015nia}
	N.~N.~Scoccola, D.~O.~Riska and M.~Rho,
	Phys.\ Rev.\ D {\bf 92} (2015) no.5,  051501
	
	\bibitem{Gerasyuta:2015djk}
	S.~M.~Gerasyuta and V.~I.~Kochkin,
	arXiv:1512.04040 [hep-ph].
	
	
	
	
	\bibitem{Bijker}
	R. Bijker et al.,  
	Eur. Phys. J. A 22, 319-329 (2004)
	
	\bibitem{PDG}
	C. Patrignani et al., (Particle Data Group), \\
	Chin. Phys. {\bf C 40}, 100001 (2016). 
	
	\bibitem{V.M.Abazov_Xi_b}
	V.M. Abazov et al. (D0 collab.), Phys. Rev. Lett. 99, 052001 (2007)
	
	\bibitem{V.M.Abazov_Omega_b}
	V.M. Abazov et al. (D0 collab.), Phys.  Rev. Lett. 101, 232002 (2008)
	
	\bibitem{Yuan:2012wz}
	S.~G.~Yuan, K.~W.~Wei, J.~He, H.~S.~Xu and B.~S.~Zou,
	Eur.\ Phys.\ J.\ A {\bf 48} (2012) 61
	
	
	\bibitem{F.Gursey}
	F. Gursey and L.A. Radicati , Phys. Rev. Lett. 13, 173 (1964)
	
	\bibitem{Cheng:2015cca} 
	H.~Y.~Cheng and C.~K.~Chua,
	Phys.\ Rev.\ D {\bf 92}, no. 9, 096009 (2015)

  \bibitem{D. symmetry breaking}
  F. Iachello,
	Nuclear Physics
        A518 (1990), 173-185
        

  
        
 
        
      \bibitem{Oh:2011}
  Y.~Oh,
  J.\ Korean\ Phys.\ Soc.\ {\bf 59}, 3344 (2011).
  
         \bibitem{Kim:2011rm}
  S.~H.~Kim, S.~i.~Nam, Y.~Oh, and H.-Ch.~Kim,
  Phys.\ Rev.\ D {\bf 84}, 114023 (2011).
  
  \bibitem{Wang:2015jsa} 
  Q.~Wang, X.~H.~Liu, and Q.~Zhao,
  Phys.\ Rev.\ D {\bf 92}, 034022 (2015).

\bibitem{Kim:2016cxr}
  S.~H.~Kim, H.~C.~Kim and A.~Hosaka,
  Phys.\ Lett.\ B {\bf 763} (2016) 358.
	
\bibitem{Camerini}
U. Camerini, J. G. Learned, R. Prepost, C. M. Spencer,
D. E. Wiser, W. W. Ash, R. L. Anderson, D. M. Ritson,
D. J. Sherden, and C. K. Sinclair, 
Phys. Rev. Lett. 35, 483 (1975).

\bibitem{Anderson}
 R. L. Anderson, Report No. SLAC-PUB-1417.
 
 \bibitem{Gittelman}
[26] B. Gittelman, K. M. Hanson, D. Larson, E. Loh, A.
Silverman, and G. Theodosiou, 
Phys. Rev. Lett. 35, 1616 (1975).
	
	\bibitem{proceedings}
Conference: C16-⁠02-⁠01.1, p.127-⁠142 Proceedings\\
e-⁠Print: arXiv:1604.02141 [hep-⁠ph]
	
	\end{thebibliography}
\end{document}